\RequirePackage{fix-cm}
\documentclass[smallextended]{svjour3}       
\smartqed  
\usepackage{graphicx}
\begin{document}
%
\title{Antimatter in the Universe : \\
Constraints from Gamma-Ray Astronomy}


\author{Peter von Ballmoos}

\institute{Institut de Recherche en Astrophysique et Plan\'etologie \\
              9, avenue du Colonel Roche, 31028 Toulouse, France \\
              Tel.: +33-561 55 66 47\\
              \email{pvb@irap.omp.eu}   }
              
\date{Received: date / Accepted: date}

\maketitle

\begin{abstract}
We review gamma-ray observations that constrain antimatter - both baryonic and leptonic - in the Universe. Antimatter can be probed through ordinary matter, with the resulting annihilation gamma-rays providing indirect evidence for its presence. 
Although it is generally accepted that equal amounts of matter and antimatter have been produced in the Big Bang, gamma-rays have so far failed to detect substantial amounts of baryonic antimatter in the Universe. Conversely, positrons are abundantly observed through their annihilation in the central regions of our Galaxy and, although a wealth of astrophysical sources are plausible, their very origin is still unknown. 
As both antimatter questions - the source of the Galactic positrons and the baryon asymmetry in the Universe - can be investigated through the low energy gamma-ray channel, the mission concept of a dedicated space telescope is sketched out. 

\end{abstract}

\section{Introduction}
\label{intro}
The present short review focuses on gamma-ray observations of baryonic and leptonic antimatter in the Universe. Following a trail that starts in our solar system, passing through our Milky Way, and out to galaxy clusters, the observational evidence (of absence) for baryonic antimatter in the Universe is reassessed, using recent high-energy data from the Fermi Gamma-ray Space Telescope ({\it Fermi}). We then revisit older Compton-GRO observations that have been used to constrain possible matter-antimatter domain boundaries in the early Universe - i.e. characteristic annihilation signatures in the diffuse Cosmic Gamma-Ray background spectrum of the MeV domain. According to the presently accepted paradigm, a matter-antimatter symmetric Universe can be ruled out on the grounds of the existing MeV observations. Today however, there is not only a need for new theoretical studies, revisiting these models in the light of 21st century cosmology, but most of all the need for a new gamma-ray mission that would be able to draw up the first map of the Cosmic Gamma Background (CGB) at MeV energies.

Positrons, on the other hand, are the most common and easily produced form of antimatter. The characteristic line at 511 keV emitted by the annihilation of Galactic positrons has been measured for almost four decades with balloon and satellite experiments. The sky map of electron-positron radiation has been drawn and redrawn by INTEGRAL, and the physical conditions in the sites where annihilation occurs are reasonably well understood. Nevertheless, the very origin of the positrons and their propagation has remained as enigmatic as ever.

Remarkably, even though their rest masses differ by nearly a factor of 2000, both the leptonic and the baryonic antimatter question can be investigated through low energy gamma-ray astronomy, at 0.5 MeV and a few MeV, respectively. In the last section, the requirements for future space-based telescope emphasizing on these questions are presented, and a possible mission-concept is outlined.

\section{Constraining baryonic antimatter with gamma-rays}
\label{sec:1}

When nucleons annihilate with anti-nucleons, they disintegrate into pions ($\pi^{0}$, $\pi^{\pm}$) which decay in flight to stable leptons and high energy photons. Characteristic gamma-rays are produced through the decay of neutral pions ($\pi^{0}$), their energies being distributed in a broad spectral bump extending from several tens of MeV to several hundreds of GeV and peaking between 100-200 MeV. A typical rest frame spectrum is shown in Fig 1 \cite{Backenstoss_1983}.

\begin{figure}
 \includegraphics[width=0.65\textwidth]{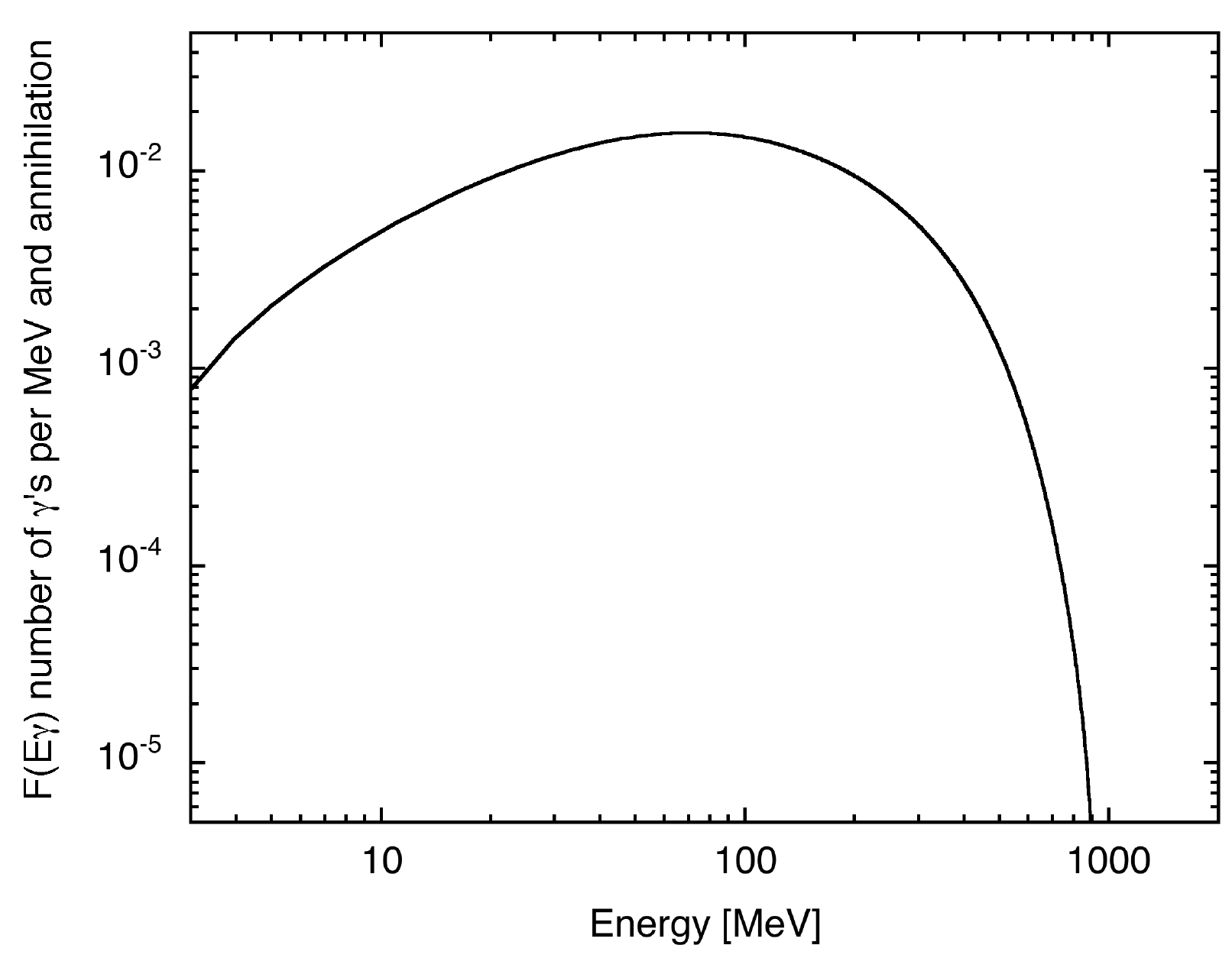}
\caption{Gamma-ray spectrum resulting from proton-antiproton annihilation at rest, according to the analytical shape function given in \cite{Backenstoss_1983}.}
\label{fig:1}       
\end{figure}

Yet, a "pion-decay bump" is not a unique signature of annihilation since it can also be created in energetic proton-proton collisions, for example.  The Fermi Large Area Telescope ({\it Fermi}-LAT) has recently detected the pion-decay feature in the spectra of two supernova remnants (SNRs), IC 443 and W44\cite{Ackermann_pion_2012}, providing direct evidence for diffusive shock acceleration of relativistic particles in these objects. Nonetheless, $\pi^{0}$ gamma-rays are fundamental for constraining the antimatter content in the Universe - i.e. setting upper limits for the antimatter fraction contained in astrophysical objects on all scales. The nature of an eventually detectable Cosmic Gamma Background (CGB) in the MeV domain is particularly important for constraining a possible Òlocally asymmetric domain cosmology (LADC).Ó 

\subsection{The "matter trail": from the solar system to clusters of galaxies}
\label{sec:2}

The fraction of nuclear antimatter in our Solar and Galactic neighborhood can be constrained by the observation of high energy gamma-rays. Following Steigman (1976) \cite{Steigman_1976} we can trace possible antibaryons by applying the same general approach to all scales : In an extended diffuse medium (such as the solar wind, galactic interstellar medium, Cosmic Rays) interacting with astrophysical bodies made from ordinary matter, the analogous antimatter bodies would necessarily interact too, producing annihilation radiation with a spectrum similar to the one shown in Fig~1.

In our solar system for example, micro-meteorites and solar wind particles continuously slam into the earth's atmosphere without causing annihilation radiation. Under the bombardment of these particles, an anti-planet would emit intense annihilation radiation. Approximating the solar wind flux at distance $d$ from the sun with $nv \approx 2 \times 10^8 (1AU/d)^{2} \ [cm^{-2} s^{-1}]$, we can estimate the annihilation flux expected from an anti-planet with radius $r$ \cite{Steigman_1976} :

%
%
\begin{equation}
F_{\gamma}(100 \ MeV) \approx 10^8\frac{r}{d^2} \ [ph \ cm^{-2} s^{-1}]
\label{e:jupiter}
\end{equation}

A hypothetical Anti-Jupiter, for example ($r= 7\times 10^7 m, d=7\times 10^{11} m$), would therefore radiate gamma-rays with a flux  $F_{\gamma}(100 \ MeV) \approx 1 \ [ph \ cm^{-2} s^{-1}]$ - eight orders of magnitude above the sensitivity limit of the {\it Fermi}-satellite.

While an antistar could in principle conceal considerable amounts of antimatter without being detectable in gamma-rays (e.g. as a dwarf and compact star), it would necessarily produce annihilation radiation during the initial and final phases of its evolution. If stellar evolution is "symmetric" to normal matter stars, an antistar would - just like a normal protostar - interact with the extended molecular cloud it is formed in, and during the final phases of its life it would bombard the surrounding interstellar medium by stellar winds (planetary nebula) and ejecta (novae, supernovae). 

Even if the signatures of the initial and final phases were neglectable, an antistar would accrete interstellar gas during its journey through the Galaxy, leading to the production of annihilation gamma-rays. We may conjecture that a population of antistars exists in the {\it unassociated} {\it Fermi} ${\gamma}$ sources. Amongst the 1 873 sources of the second {\it Fermi}-LAT point-source catalog (2FGL) \cite{Nolan_2012} about a third are lacking reliable association with sources detected in other wavelength bands -- i.e. 576 sources. For a conservative estimate of an "antistar-density",  unassociated  sources (100 MeV $< E_{\gamma} <$ 1 GeV) were selected, with fluxes $\phi_{\gamma}\ge 10^{-7}$ ph cm$^{-2}$ s$^{-1}$. If those sources  ($N_{\bar{*}}$=75 objects) were to radiate through Bondi Hoyle accretion of interstellar galactic gas onto their surfaces ($L_{\gamma} \sim 3 \cdot 10^{35}(M/M_{sun})^2 v^{-3}_6$, with $v_6$ being the mean stellar velocity relative to the gas, see \cite{Steigman_1976}) they would have to be situated closer than 150 pc from the Sun. On the other hand, about $N_{*} \simeq 2\cdot 10^6$ normal stars are expected within this distance. We therefore must be living in a local Galactic neighborhood that is matter/antimatter asymmetric with an $f_* = N_{\bar{*}} / N_{*} < 4 \cdot 10^{-5}$.

For our Milky Way, the Magellanic Clouds, or the Andromeda galaxy, upper limits to an antimatter fraction $f_{ISM}$ in the  interstellar medium (ISM) can be assessed by requiring that the emissivity  $q_{ann}$ resulting from $\bar{p}p$ annihilation must necessarily be inferior to $q_{\gamma}$, the observed $\gamma$-ray emissivity  \textgreater 100 MeV. We note, moreover, that $q_{\gamma}$ is well described by cosmic-ray interaction with the gas. With roughly 3 $\gamma$'s emitted per $\bar{p}p$ annihilation, the annihilation emissivity per hydrogen atom is $q_{ann} \simeq 3 \cdot f_{ISM} \cdot  n_H \cdot  \sigma_{ann} \cdot v$, where $n_H \simeq$ 1 cm$^{-3}$ is the average density of the ISM, and $\sigma_{ann} v \simeq 10^{-10}$ cm$^3$ s$^{-1}$ a lower limit to the annihilation rate coefficient for low temperature (T$\leq 10^4$ K) $\bar{H}H$ annihilation  \cite{Steigman_1976}. In table 1  conservative upper limits to the antimatter fraction are estimated by
%
%
\begin{equation}
f_{ISM} < \frac{q_{\gamma}}{3 \cdot \ n_{H}\cdot \sigma_{ann} v} ,
\label{e:galaxies}
\end{equation}

with the average integrated \textgreater 100 MeV gamma-ray emissivity per H-atom $q_{\gamma}$ derived from recent {\it Fermi}-LAT observations \cite{Abdo_M31_2010}.

\begin{table}
\caption{Upper limits to the antimatter fraction $f_{ISM}$ in the interstellar medium (ISM) of our Milky Way and closeby galaxies, based on the observed gamma-ray emissivity per hydrogen atom $q_{\gamma}$, derived from recent {\it Fermi}-LAT observations (Abdo et al. 2010 \cite{Abdo_M31_2010} and references therein).}
\label{tab:1}       
\begin{tabular}{lccc}
\hline\noalign{\smallskip}
   & ${L_{\gamma}} $ ({\it Fermi}) & $q_{\gamma} $ & $f_{ISM}$  \\
  \noalign{\smallskip}
   & $10^{41}$ ph s$^{-1}$  &  $10^{-25}$ ph s$^{-1}$   H-atom$^{-1}$& $10^{-16} $  \\
\noalign{\smallskip}\hline\noalign{\smallskip}
Milky Way & 11.8 $\pm$ 3.4 & 2.0 $\pm$  0.6 & 8.6\\
Andromeda & 6.6 $\pm$  1.4 & 0.7 $\pm$  0.3 &3.3\\
Large Magellanic Cloud & 0.78 $\pm$  0.08 & 1.2 $\pm$  0.1 & 4.3 \\
Small Magellanic Cloud & 0.16 $\pm$  0.04 & 0.31 $\pm$ 0.07 & 1.2\\
\noalign{\smallskip}\hline
\end{tabular}
\end{table}

On an even larger scale, the fraction of antimatter contained in galaxy clusters can be constrained by matching up the measured soft X-ray flux from the hot intracluster medium with upper limits on the high energy gamma flux from the clusters. We again follow the argumentation of Steigman (1976) \cite{Steigman_1976}, however using recent {\it Fermi} data. The bulk of the baryonic matter contained in galaxy clusters is in a diffuse, hot, metal-enriched plasma that radiates primarily in the soft X-ray domain. If this intracluster medium were composed of a blend of matter and antimatter,  the same binary Coulomb collisions that produce the soft X-rays through  thermal bremsstrahlung would inevitably generate high energy $\gamma$-rays through $p\bar{p}$ annihilation. The antimatter fraction $f_c$ can be constrained to

%
%
\begin{equation}
f_c \  \le \ 3.0\times 10^{-11} \ T_{keV}  \ \frac{F_{\gamma}}{F_x} 
\label{cluster}
\end{equation}

where $T_{keV}$ is the temperature of the intracluster gaz in [keV],  $F_X$  the X-ray flux in the 2--10 keV band [erg cm$^{-2}$ s$^{-1}$], and $F_{\gamma}$ the upper limit to the gamma ray flux above 100 MeV. Figure 2 compares X-ray data from \cite{Edge_1990} for a number of galaxy clusters with recent gamma-ray upper limits from the {\it Fermi}-LAT telescope \cite{Ackermann_clusters_2010} (with integral fluxes  extrapolated down to $\ge$ 100 MeV). As {\it Fermi} features sensitivity improvements  of more than an order of magnitude over the EGRET data used in \cite{Steigman_2008}, the antimatter fractions $f_c \simeq 10^{-8}$ obtained here are the lowest upper bounds to the antimatter contained at  large scales. Although the bullet cluster "only" has a $f_c \sim 5 \times 10^{-7}$, its case is of particular interest as it consists of two colliding clusters of galaxies that were initially separated by $\sim$ 20 Mpc. Further observations will have to confirm the absence of matter-antimatter blends on scales of up to tens of Mpc. We already can conclude that our local Universe -- at scales ranging from AU's to Mpc's -- contains predominately matter and very little antimatter.

\begin{figure}
 \includegraphics[width=0.75\textwidth]{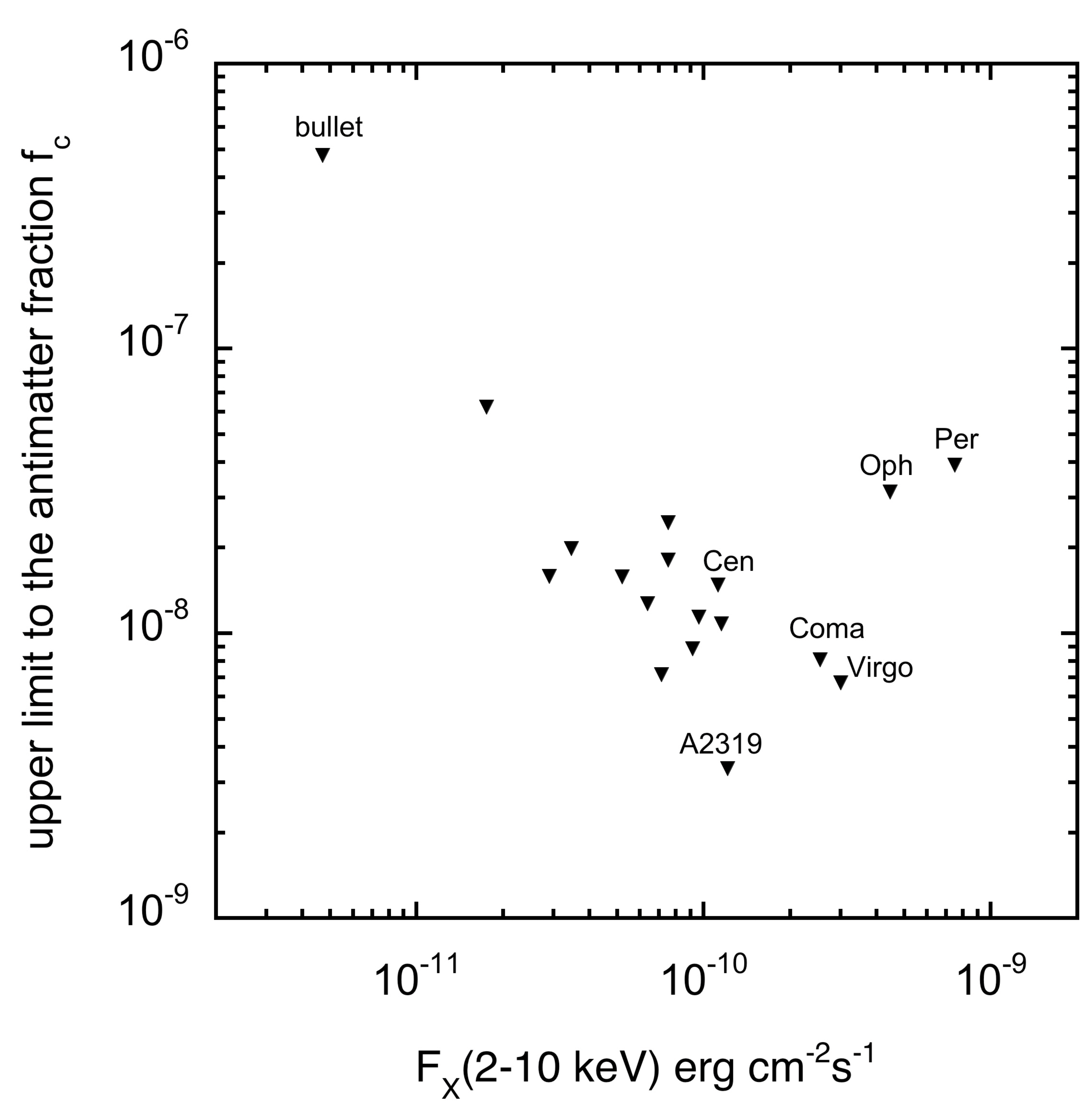}
\caption{Constraining the antimatter fraction $f_c$ in clusters of galaxies : recent gamma-ray upper limits from the {\it Fermi}-LAT telescope \cite{Ackermann_clusters_2010} are compared with the X-ray flux (2-10 keV) \cite{Edge_1990} of 18 galaxy clusters.}
\label{fig:2}       
\end{figure}
%

\subsection{Antimatter at high redshifts}
\label{sec:4}

The presence of putative matter-antimatter domain boundaries on the largest scales of the observable Universe can be constrained by studying the Cosmic Gamma Background (CGB) in the MeV domain. Assuming a Universe divided into equal matter- and antimatter regions with annihilations annihilations taking place on the domain boundaries, Stecker et al. (1971) \cite{Stecker_etal_1971} were the first to derive a spectrum for the CGB by solving a cosmological photon-transport equation. After having undergone absorption, scattering, and a cosmological redshifing, the source $\gamma$-ray function -- a $\pi^{0}$ rest frame spectrum peaking at 68 MeV -- is transposed to the key of low energy $\gamma$-rays, resulting in an "MeV bump". The MeV feature is essentially due to a z $\sim$ 100 redshift of the $\pi^{0}$-decay bump falling off to higher energies in an approximate power law; below $\sim$1 MeV the spectrum is flattening out due to energy loss from pair-production and Compton-scattering at high redshifts (z \textgreater 100). 

In the seventies, gamma-ray spectrometers flown on the Apollo flights \cite{Trombka_1977} and balloon-borne Compton telescopes  \cite{Schonfelder_1980}, \cite{White_1977} have indeed measured an MeV feature in the diffuse $\gamma$-ray background, bolstering ideas of cosmic antimatter domain boundaries. In the nineties however, after carefully studying the cosmic $\gamma$-ray background (and most importantly, the predominant instrumental background) for nine years, the COMPTEL telescope on the Compton Gamma Ray Observatory could not confirm the "MeV bump". Instead, a featureless CGB power spectrum was observed by COMPTEL (Fig. 3). What's more, the measured spectra of resolved Active Galactic Nuclei (AGN) suggest that the remaining CGB may likely be explained as a superposition of distant unresolved AGNs.

\begin{figure}
 \includegraphics[width=0.9\textwidth]{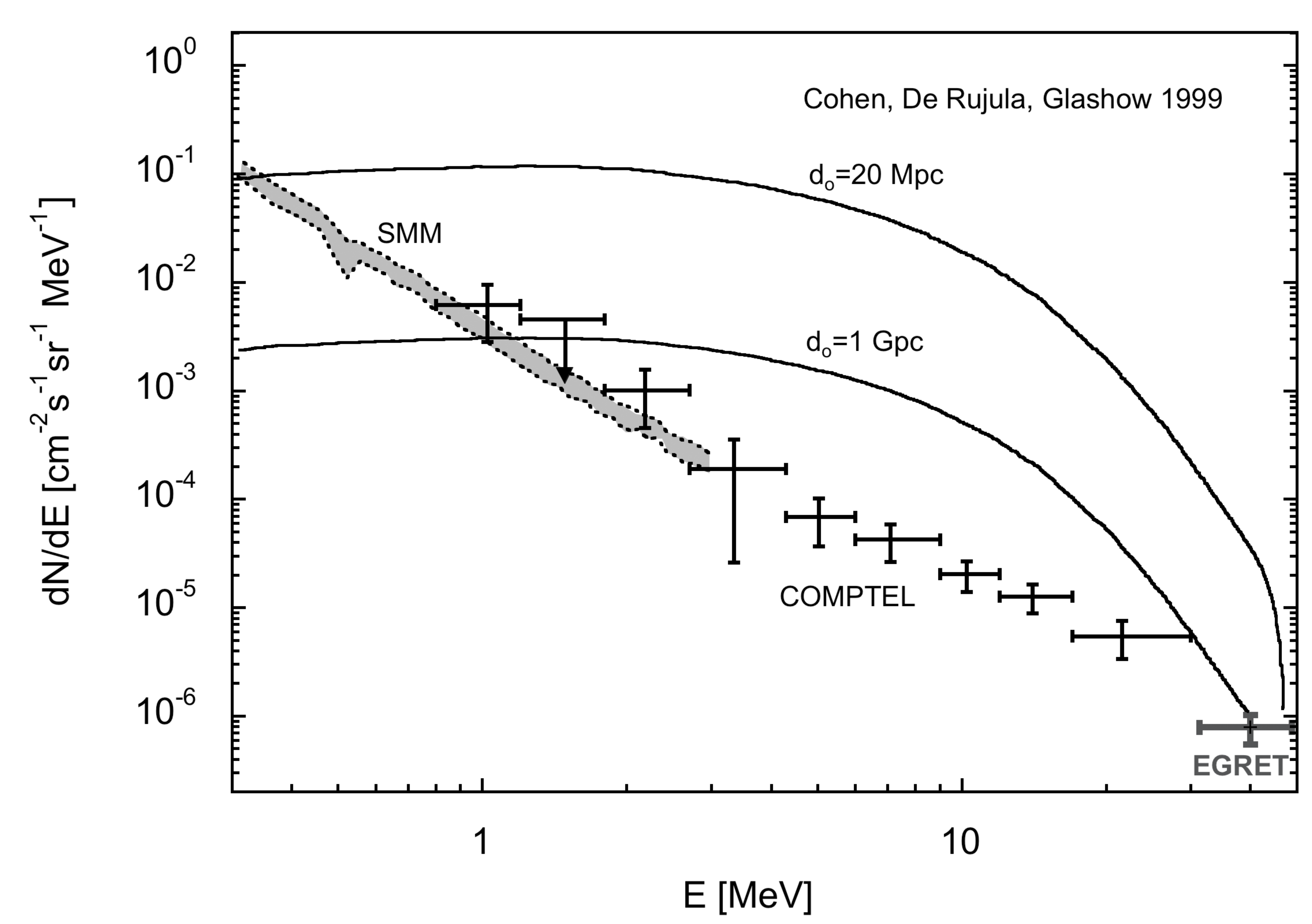}
\caption{Model spectra for a matter-antimatter symmetric Universe by Cohen, De R\`ujula, and Glashow (1998) \cite{CDG_1998} compared with $\gamma$-ray observations in the relevant MeV range : CGRO/COMPTEL data are from Weidenspontner and Varendorff (2001) \cite{Weidens_2001}, the best fit model for the SMM data \cite {Watanabe_1997} is shown by the grey area between dotted lines is delimiting the range of uncertainties, and the CGRO/EGRET datapoint is from \cite{Strong_2004}}
\label{fig:3}       
\end{figure}

When combined with the quality of the observed Cosmic Microwave Background (CMB), the CGB measured by COMPTEL was found to make a remarkably strong argument against a baryon-symmetric Univers (Cohen, De R\`ujula, and Glashow, 1998 : CDG98 \cite{CDG_1998}). The minute temperature fluctuations in the CMB (parts in $10^{-5}$) correspond to a very uniform matter density at recombination, implying that matter and antimatter must necessarily have come into contact at this epoch. Annihilation would therefore have been unavoidable in the early Universe. Cohen et al. \cite{CDG_1998} develop a sophisticated scenario for baryonic annihilation radiation undergoing scattering and cosmological redshift, integrating the annihilation at the domain interface from recombination (z$\sim$1100) to the onset off structure formation (z$\sim$20). Figure 3 compares the  calculated model spectra for domains with comoving sizes $d_o$=20 Mpc and $d_o$=1 Gpc with the CGB as measured by COMPTEL. In 1998, Cohen, De R\`ujula, and Glashow thus assert that "a matterÐ-antimatter symmetric Universe is empirically excluded".

Today, fifteen years after the epochal CDG98 paper, the question of baryonic antimatter in the Universe certainly merits a new look. First, the cosmological input parameters should be tuned from $\rm \Omega_m$=1, $\rm \Omega_{\Lambda}$=0, h=0.75, to the parameters of the "Planck era", i.e. $\rm \Omega_m$=0.32, $\rm \Omega_{\Lambda}$=0.68, h=0.67.8. At the same time, some of the hypothesis used in CDG98 could be reconsidered, and processes that were neglected at the time should be included : e.g. the consequences of annihilation in the pre-recombination area on the CGB, or the effect of magnetic fields parallel to the domain boundaries. The latter would probably affect the diffusion of particles across the boundary an hence change the annihilation rate \cite{Stecker_2003}.

With Stecker (2003) \cite{Stecker_2003}, we conclude that "absence of evidence is not necessarily evidence of absence". Ultimately only a new gamma-ray mission can clarify the observational situation. At the least, a new look at the Cosmic Gamma Background will have to show how strong the asymmetry is, establishing a filling factor - or antimatter fraction $f_{CGB}$ for distance scales beyond Galaxy Clusters - i.e tens of Megaparsecs.

\section{Leptonic Antimatter - Positron Annihilation}
\label{sec:2}

Since Anderson's discovery of the positron in 1932, the question of its existence in the Universe - and of antimatter in general - has puzzled astrophysicists. As positron production is ubiquitous in a large number of high-energy processes, identifying possible primordial antimatter through its e$^{-}$e$^{+}$ annihilation seems out of the question, though this may look tempting because of the clear annihilation feature at 511 keV. Besides the positrons abundantly produced in radioactive decays and by cosmic rays in the atmosphere -- and our telescopes --  it was soon supposed that they might be generated in a multitude of astrophysical environments (nucleosynthesis, neutron stars, pair plasma etc.). 

Line emission at 511 keV from the Galactic Center region has first been observed in the early seventies with balloon experiments \cite{Johnson:1972} \cite{Albernhe:1981}, \cite{Leventhal:1978}. The eighties were marked by ups and downs in the measured 511\,keV flux by Ge-detectors flown on stratospheric balloons and on the HEAO-3 satellite. The variable results were first interpreted as the signature of a~compact source of annihilation radiation at the Galactic Center (see e.g. Leventhal, 1991~\cite{Leventhal:1991}. Yet in 1990, neither the eight years of SMM data~\cite{Share:1990}, nor the revisited data of the HEAO-3 Ge detectors~\cite{Mahoney:1993}, showed evidence for variability in the 511\,keV flux. In the nineties, CGRO-OSSE measured steady fluxes from a~galactic bulge and disk component and rough skymaps of the Galactic Center region \cite{Purcell:1997} became available based on data from OSSE, SMM and TGRS.

\begin{figure}
\includegraphics[width=0.7\textwidth]{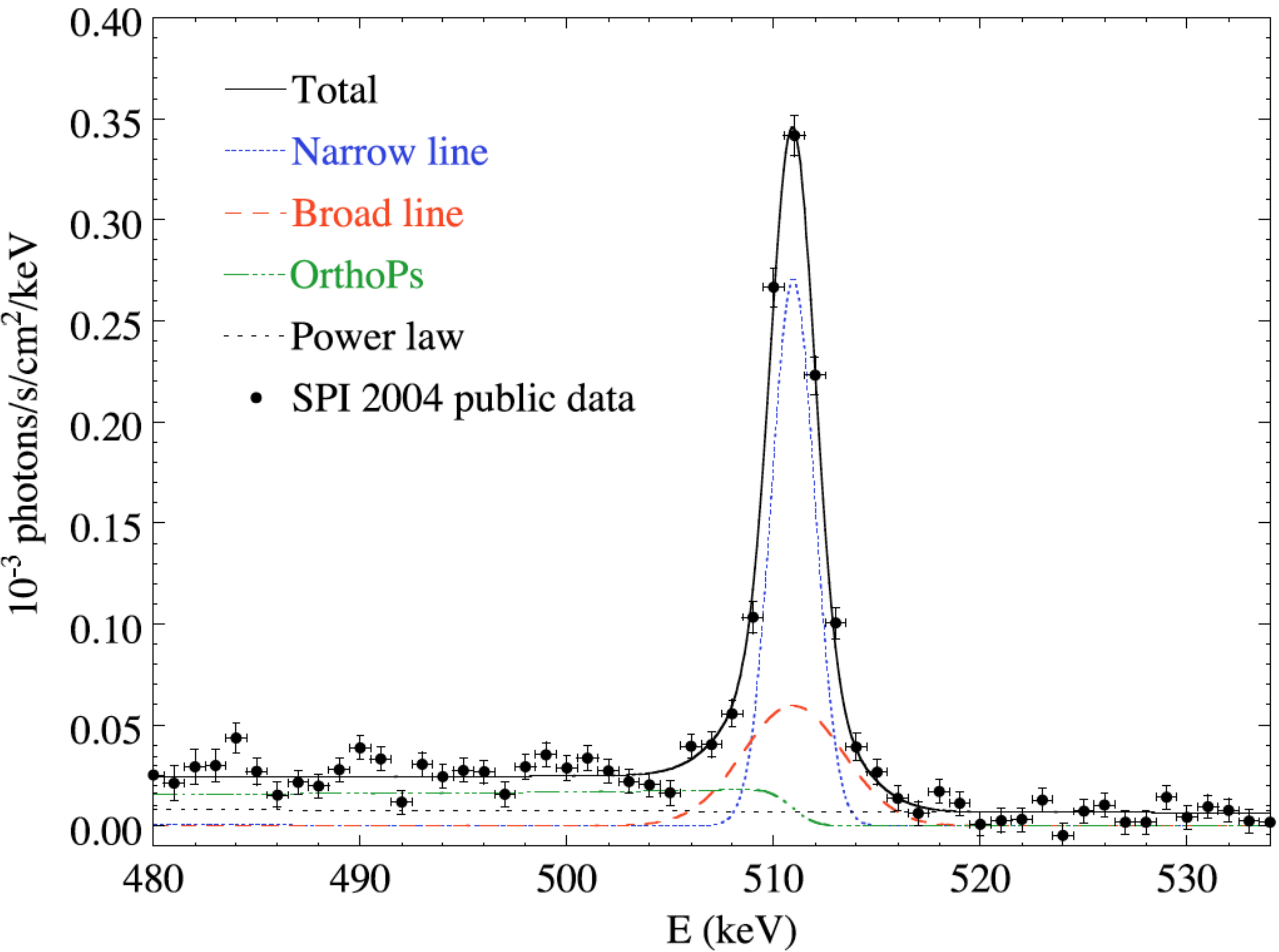}
\caption{The spectrum of the Galactic center e$^{-}$e$^{+}$ annihilation observed by INTEGRAL/SPI (Jean et al.2006). The fit of the spectrum  -- i.e. the narrow and a broad Gaussian line -- is constraining the physical conditions in the sites where annihilation occurs, i.e. the neutral and ionized warm phases of the interstellar medium \cite{Jean_2006}.}
\label{fig:4}       
\end{figure}

With the INTEGRAL observatory \cite{Vedrenne_2003} that was launched in October 2002 -- and still is operational as of today -- our knowledge in positron astrophysics has significantly improved: using data from its spectro-imager SPI, a first {\it all}-sky map of electron-positron radiation has been drawn and the physical conditions of the medium in which the positrons annihilate have been constrained by spectroscopy :

The main spectral features of the Galactic e$^{-}$e$^{+}$ emission measured by INTEGRAL/SPI are a narrow and broad 511 keV line plus an orthopositronium continuum \cite{Jean_2006} (Fig. 4), indicating that most positrons annihilate after positronium formation (positronium fraction of $f_{Ps}=97\pm 2 \%$). The spectrum is interpreted as the signature of positrons annihilating in the warm neutral phase and the warm ionized phase of the interstellar medium. Although spectroscopy characterizes the physical conditions in which the positrons perish, it can't help us with finding their birthplace. The origin of the positrons may be constrained by the morphology of the annihilation radiation, provided that the positrons do not travel too far before annihilating. The sky map displayed in Fig. 5 represents a best-fit model based on ten years of SPI data \cite{Skinner_2013}. The 511 keV emission is strongly concentrated in a ÔbulgeÕ region of $\sim 6^{\circ} $ FWHM diameter in the direction of the Galactic Center. There is evidence for a slight offset (0.5$^\circ$) of the bulge to negative Galactic longitudes.   The Galactic disk is detected with a much lower surface brightness, implying an unexpectedly high bulge-to-disk ratio. The total Galactic e$^{+}$ annihilation rate is of the order of $2 \times 10^{43}\ e^{+} s^{-1}$. No evidence has been found for point sources, a "galactic fountain", or from source regions other than our galactic disk and bulge. 

While the origin of the positrons annihilating in the disk could readily be explained by several types of astrophysical objects -- the most plausible source being radioactive isotopes ($^{56}$Ni, $^{44}$Ti, $^{26}$Al) produced by supernovae - the Galactic Bulge positrons do not fit any 'classic' source distribution. Besides non-stationary explanations such as a recent starburst episode close the Galactic Center black hole Sag A* \cite{Alexis_2013}, cusp shaped distributions of dark matter distributions could fit the bulge positrons. This latter hypothesis recently has gotten renewed interest as certain dark matter halo models \cite{Kuhlen_2013} invoke an offset in the central density peak towards negative Galactic longitudes ...

INTEGRAL's achievements have permitted us to better understand the places and conditions where galactic positrons annihilate. However, the production, the transport and the annihilation of the positrons has been -- and remains -- one of the key questions in $\gamma$-ray astronomy.  

\begin{figure}
 \includegraphics[width=0.75\textwidth]{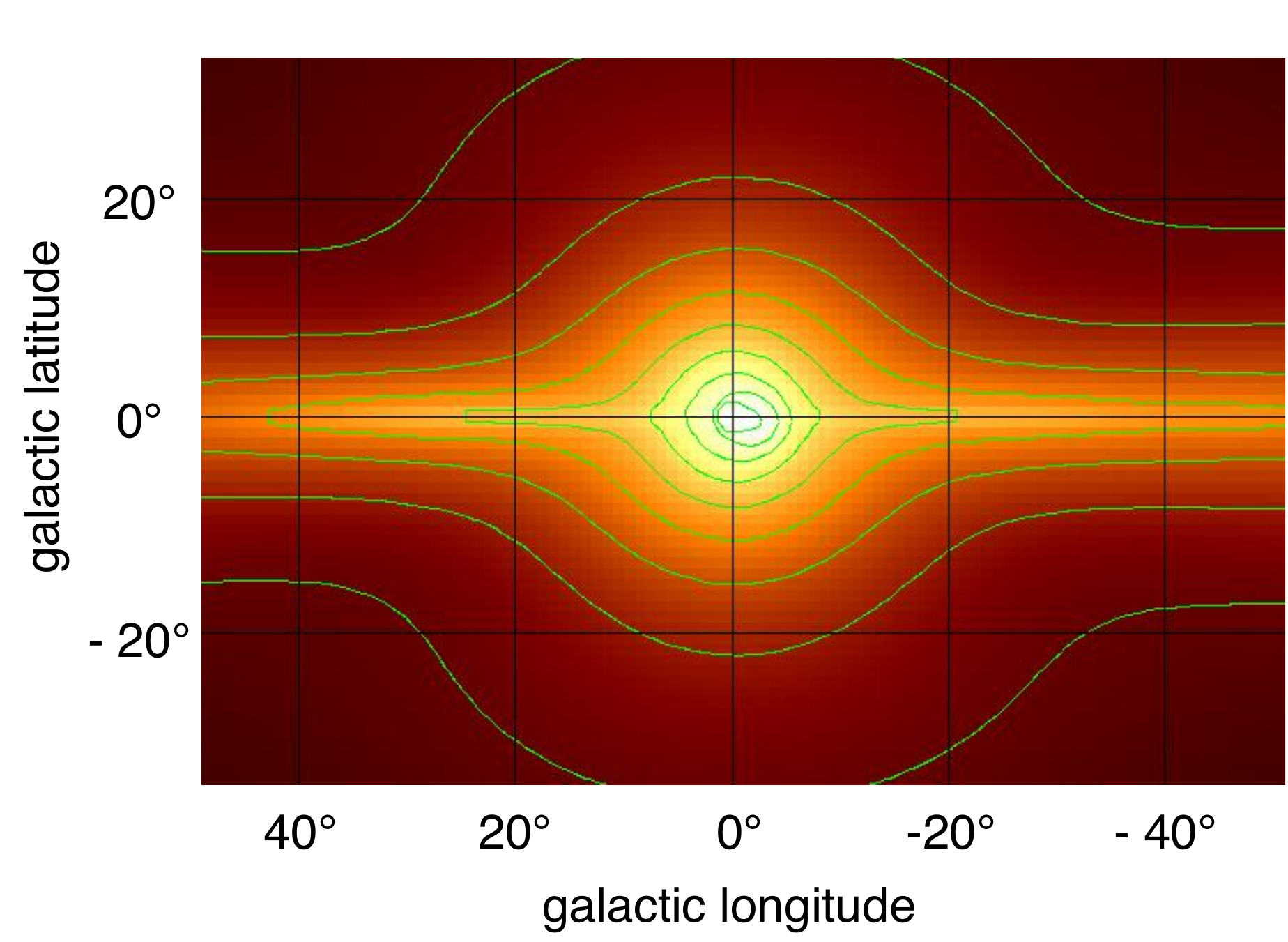}
\caption{Best-fit model of the 511-keV e$^{-}$e$^{+}$ annihilation line in the Galactic Center region \cite{Skinner_2013}. The bright central bulge is by far the most prominent feature. The map is based on observations with the imaging spectrometer SPI on board the INTEGRAL satellite, and uses 10 years of mission data.}
\label{fig:4}       
\end{figure}
%

\section{Conclusion : a $\gamma$-ray telescope for cosmic antimatter studies}
\label{sec:3}

Gamma-ray observations in the 100 MeV domain (CGRO-EGRET, {\it Fermi}-LAT) have virtually excluded the existence of substantial amounts of "local" baryonic antimatter, constraining its fraction to a few times $10^{-5} - 10^{-9}$ in structures as large as galaxy clusters.  Likewise, a universal matter-antimatter symmetric (B=0) Universe may already be excluded by theory (e.g. CDG98  \cite{CDG_1998}) on the grounds of existing COMPTEL data.  However, the present data do not constrain the cosmological antimatter fraction or filling factor - or very weakly at best. None of the existing data allowed to study anisotropies or possible structure on the Cosmic Gamma-ray Background. 

To clarify the experimental situation in the MeV range, Stecker (2003) \cite{Stecker_2003} recommends the construction of "a dedicated MeV background satellite detector experiment designed to be as clean from radiation induced intrinsic contamination as possible". In passing, we note that there also is a need for new theoretical studies, revisiting the models of cosmological antimatter in the light of 21$^{st}$ century cosmology.

Constraining the antimatter content in the early Universe is one of the fundamental questions of physics only observational MeV-astronomy will ultimately be able to tackle \cite{CDG_1998}, \cite{Stecker_2003}. A new look at the Cosmic Gamma-Ray Background (CGB) with a sensitivity well beyond of what was possible in the 1990's will not only constrain the filling factor of baryonic antimatter, but it would allow to search for possible structures in the CGB. As the putative domain boundaries create "ribbons" in the sky, the Cosmic Gamma-Ray Background should eventually be analyzed in the same way the Cosmic Microwave Background is studied (component separation of possible anisotropies, multipole analysis). Assessing the CGB is admittedly a highly complex task, it requires exhaustive modeling of the instrumental background, and a solid knowledge of foreground emissions.

Needless to say, these foreground emissions are by themselves of paramount astrophysical interest - they comprise the vast variety of astrophysical objects presently studied by high-energy astronomers. A wide-field telescope with the ambitious goal of mapping the Cosmic Gamma-Ray Background will necessarily measure a large number of astrophysical objects.

With respect to "foreground" astrophysics, the capability of measuring polarization would provide a powerful new diagnostic of magnetic fields for understanding the nature of the central machine in $\gamma$-ray bursts, for studying acceleration in neutron star magnetospheres and for their role in the origin of $\gamma$-ray emission and jets from accreting black holes, both galactic and extragalactic. Such a telescope would simultaneously perform a detailed all-sky survey of the radioactive Milky Way, clarifying long-lived activities from supernovae and novae; and studying excitation lines from as yet unknown MeV cosmic-rays interacting with the interstellar medium. A deep all-sky survey naturally monitors galactic black holes, neutron stars, pulsars, magnetars, and simultaneously, a full sky of Active Galactic Nuclei. Last but not least, a future mission in this domain must significantly deepen the study of the galactic 511 keV emission, and ultimately unveil the origin of galactic positrons. While there is a long list of potential sources for the MeV positrons - including supernovae, micro-quasars, x-ray binaries, gamma-ray bursts, and annihilation or decay of hypothetical light dark matter particles - their origin and their propagation has remained as enigmatic as ever.

A future space-based telescope emphasizing on baryonic and leptonic antimatter - besides all the above topics - will have to fulfill specific requirements: the entire sky is to be surveyed in the energy range from 0.1 to at least 20 MeV, featuring a sensitivity at least one order of magnitude better than previous missions. As virtually every background photon has to be accounted for to extract a CGB, an ultra-low instrumental background -- well controlled and understood -- is probably another foremost requirement for such a telescope.

The {\it All-Sky Compton Imager} (ASCI) proposed in \cite{DUAL_2012} and \cite{ASCI_2014} is a mission concept particularly well suited to the task of measuring the Cosmic Gamma-Ray Background -- and simultaneously covering the wide range of science topics in gamma-ray astronomy. ASCI features a compact array of cross-strip germanium detectors (mass $<$  100 kg), a concept that has  been successfully tested during stratospheric balloon flights \cite{Bandstra_2011}. The omindirectional detector is situated on a deployable mast at a distance of 10 m from the spacecraft which is orbiting around the second Lagrange point of the Sun-Earth system, L2. The instrument therefore not only avoids Earth albedo- and spacecraft-induced background, but it also benefits from a continuous all-sky exposure~: every single source in the sky is observed during the entire mission lifetime ($T_{life}\simeq 10^8 sec$). With its high spectral and 3-D spatial resolution, the ASCI will perform sensitive $\gamma$-ray spectroscopy and polarimetry in the energy band 100 keV-20 MeV. Despite its modest size, ASCI's $\gamma$-ray line sensitivity after its nominal lifetime of 3 years  is $\sim 10^{-6} \ ph \ cm^{-2} s^{-1}$ at 1 MeV for every $\gamma$-ray source in the sky !
          
It is noteworthy that both antimatter questions - the origin the Galactic positrons and the baryon asymmetry - await investigation through low energy gamma-ray astronomy, at 0.5 MeV and a few MeV, respectively. We therefore stress the importance of a dedicated MeV gamma-ray mission emphasizing on cosmic antimatter studies.



\end{document}